# RHESSI LINE AND CONTINUUM OBSERVATIONS OF SUPER-HOT FLARE PLASMA

A. CASPI[1] AND R. P. LIN[1,2]

[1]Space Sciences Laboratory and Department of Physics, University of California, Berkeley, CA, 94720-7450
[2]School of Space Research, Kyung Hee University, Republic of Korea

## ABSTRACT

We use *RHESSI* high-resolution imaging and spectroscopy observations from ~5 to 100 keV to characterize the hot thermal plasma during the 2002 July 23 X4.8 flare. These measurements of the steeply falling thermal X-ray continuum are well fit throughout the flare by two distinct isothermal components: a super-hot ($T_e > 30$ MK) component that peaks at ~44 MK and a lower-altitude hot ($T_e \lesssim 25$ MK) component whose temperature and emission measure closely track those derived from *GOES* measurements. The two components appear to be spatially distinct, and their evolution suggests that the super-hot plasma originates in the corona, while the *GOES* plasma results from chromospheric evaporation. Throughout the flare, the measured fluxes and ratio of the Fe and Fe–Ni excitation line complexes at ~6.7 and ~8 keV show a close dependence on the super-hot continuum temperature. During the pre-impulsive phase, when the coronal thermal and non-thermal continua overlap both spectrally and spatially, we use this relationship to obtain limits on the thermal and non-thermal emission.

## 1. INTRODUCTION

Observations of the soft X-ray (SXR; ~0.1–10 keV) thermal continuum and those of thermally-excited ion emission lines — primarily from Fe XXIV and XXV — have shown that hot, ~10–20 MK, thermal plasma is present in nearly all solar flares. At hard X-ray (HXR) energies (≳20 keV), flare spectra generally fit power laws (Kane *et al.* 1980), consistent with bremsstrahlung emission from accelerated (non-thermal) electrons colliding with the ambient atmosphere. Early HXR observations, however, had coarse resolution ($\Delta E/E$ of ~25% to ~133%), so emission from very hot ($T_e \gtrsim 100$–1000 MK) thermal plasma could not be ruled out (e.g., Crannell *et al.* 1978). The flare SXR flux-versus-time profile is often observed to be proportional to the time integral of the HXR or microwave flux (the "Neupert effect" — Neupert 1968; Dennis & Zarro 1993), indicating that the hot plasma that fills the SXR-emitting loops may be heated by collisions of flare-accelerated electrons with the chromosphere, as evidenced by measurements of blueshifted line profiles consistent with evaporation of heated chromospheric material from loop footpoints (Antonucci 1989).

The first high-resolution (~2 keV FWHM) measurement of HXR (~13–300 keV) flare spectra (Lin *et al.* 1981), using cryogenically cooled germanium detectors (GeDs), showed that the spectrum above ~33 keV was a double power law with a sharp break, inconsistent with a thermal source. Below ~33 keV, however, the measurements resolved a steeply falling thermal continuum with plasma temperatures of up to ~34 MK, much hotter than previously observed. The thermal spectra were precise enough that a stringent mathematical test could be applied to confirm their thermal origins (Emslie *et al.* 1989). Subsequent continuum and Fe XXVI line observations by *Hinotori* (Tanaka 1987) and *Yohkoh* (Pike *et al.* 1996) showed that these super-hot ($T_e \gtrsim 30$ MK) plasmas are common in *GOES* X-class flares, but their origins remain poorly understood.

The *Reuven Ramaty High Energy Solar Spectroscopic Imager* (*RHESSI*; Lin *et al.* 2002) provides high spectral and spatial resolution X-ray observations down to ~3 keV, enabling measurements of the thermal continuum from plasmas with temperatures down to ≲10 MK and of the fluxes in the Fe (~6.7 keV) and Fe–Ni (~8 keV) excitation line complexes (Phillips 2004). The ratio of the flux in the Fe complex to that in the Fe–Ni complex provides an independent measure of temperatures from ~15 MK (where the Fe–Ni line becomes detectable above the thermal continuum) up to ~60 MK (above which the ratio is relatively insensitive), with little dependence on the elemental abundances (since Ni contributes ≲20% of the Fe–Ni flux and since Fe and Ni have similar first-ionization potentials; K. Phillips 2005, private communication).

Here, we use *RHESSI* imaging and spectroscopy to determine the location and morphology of the thermal and non-thermal sources, and to obtain the line fluxes, continuum temperatures, and emission measures of the thermal plasmas throughout the 2002 July 23 X4.8 flare. The spectra and images indicate that there are two spatially and spectrally separate isothermal populations throughout the flare: a super-hot (~21–44 MK) component and a hot (~13–24 MK) component that closely tracks the temperature and emission measure derived from *GOES* measurements. The presence of the super-hot component during the flare pre-impulsive phase, when little or no footpoint HXR emission is detected, suggests that it originates in the corona and not through chromospheric evaporation.

## 2. OBSERVATIONAL DETAILS

The front segments of *RHESSI*'s GeDs provide ~1 keV FWHM spectral resolution, capable of resolving the steeply falling (*e*-folding of ~2 keV) super-hot continuum, while *RHESSI*'s imaging spectroscopy allows characterization of thermal and non-thermal sources with angular resolution down to ~2 arcsec (see Lin *et al.* 2002, and references therein).

The 2002 July 23 X4.8 flare (Figure 1) divides naturally into a pre-impulsive phase (~00:18–00:26 UT) dominated by an HXR coronal source, an intense impulsive phase (~00:26–00:43 UT), and a decay phase (≳00:43 UT) dominated by slowly decreasing SXR emission (Lin *et al.* 2003). The entire flare was observed through the thin or thick+thin aluminum attenuator disks that reduce the intense incident SXR flux to minimize detector deadtime (Smith *et al.* 2002). Transitions between attenuator states were used to obtain precise intercalibration of the relative attenuator responses and to optimize the software correction of pulse pile-up (see Caspi 2010 for details). The calibration of the thin attenuator response was verified for flares also observed by the Solar X-Ray Spectrometer (SOXS) silicon PIN instrument (Jain *et al.* 2005), where *RHESSI* and SOXS 6–12 keV spectra agreed to within ~5%–10%. The background due to escaping Ge K-shell fluorescence photons was calibrated from the constant minimum ratio of ~4–5 keV counts to ~14–15 keV incident photons typically reached during the flare. In either attenuator state, low-energy photons were predominantly detected at the GeD center where charge collection is optimal, resulting in a ~0.75 keV FWHM resolution (determined by fits to the Fe line complex) for the best GeD, used for spectral analysis with the `Object Spectral Executive (OSPEX)` package[3] in the `SolarSoft` IDL software suite[4].

The fluxes in the Fe and Fe–Ni line complexes, and their ratio, were determined by subtracting the underlying ~5–10 keV continuum — well fit by a power law — and modeling the line complexes as Gaussian functions with intrinsic FWHMs of 0.15 keV (approximately the span of the prominent individual lines at ~20–50 MK) centered at their mean energies of 6.680 and 8.015 keV, respectively (Phillips 2004).

The continuum above ~10 keV was initially fit with a photon emission model consisting of a super-hot isothermal and a non-thermal component, convolved with the instrument response. The isothermal was modeled using the CHIANTI spectral code (ver. 5.2; Landi *et al.* 2006) with coronal abundances:

$$I_{th}(E,T_e,Q) \propto n_e^2 V g(E,T_e) \frac{e^{-E/k_B T_e}}{E\sqrt{T_e}} \text{ (photons s}^{-1}\text{ cm}^{-1}\text{ keV}^{-1}\text{)}$$

with fit parameters of electron temperature $T_e$ and volume emission measure $Q = n_e^2 V$, where $E$ is photon energy, $V$ is the thermal source volume, $n_e$ is the electron number density (assumed uniform with unity filling factor), and $g$ is the Gaunt factor including contributions from both free-free (bremsstrahlung) and free-bound (radiative recombination) interactions. The non-thermal continuum was modeled by a power law (or double power law where needed) with a low-energy electron cutoff, typically constrained only as an upper bound since thermal emission dominates at lower energies. The free model parameters (isothermal emission measure, temperature; non-thermal flux, power-law exponent, cutoff [plus second exponent and break energy where needed]) were optimized by iterative chi-squared minimization using `OSPEX` with the systematic uncertainty parameter set to 0%.

Subtracting the best-fit model from the observations revealed a significant residual continuum below ~15 keV that decreases rapidly above ~10 keV and appears to fit well to a second hot (but cooler) isothermal. We therefore refit the entire ~4.67–100 keV continuum — excluding the lines — with a revised model (Figure 2) including both super-hot and hot isothermals (Figures 1(c) and (d)) plus a non-thermal power law throughout the impulsive and decay phases, yielding reduced $\chi^2$ values of ~0.7 to ~2.4 (averaging ~1.4) with no significant remaining continuum.

The Fe and Fe–Ni line complex fluxes and their ratio show a correlation with the super-hot continuum temperature (Figure 3), with a steeper functional dependence at lower temperatures (≲25 MK). CHIANTI predictions of the line fluxes and ratio for the two thermal plasmas combined agree qualitatively with the observations, but quantitatively are significantly larger — by, on average, ~55%, ~20%, and ~34%, respectively, with larger deviations at lower temperatures (Caspi 2010). Throughout the pre-impulsive phase, when a non-thermal HXR source is observed in the corona (initially with no detectable footpoint emission) cospatial with a thermal source, the continuum spectrum can be fit by a wide range of model parameters (cf. Holman *et al.* 2003). By assuming that the empirical correlation between the Fe and Fe–Ni lines and the super-hot continuum observed during the rest of the flare (Figure 3) also holds here, we obtain constraints on the super-hot temperature and emission measure during this phase.

The thermal source size was determined from the 50% intensity contour of (thermally dominated) 6.2–8.5 keV *RHESSI* images using Clean with uniform weighting (Hurford *et al.* 2002). Simulations for elliptical Gaussian sources show that the length $2a$ and width $2b$ — corrected for broadening by the point-spread function — are determined to ~7%, yielding a ~23% uncertainty in the ellipsoidal volume $V = (4/3)\pi ab^2$. We arbitrarily assume a volume of $V/2$ each for the super-hot and hot plasmas and derive their thermal electron densities $n_e = \sqrt{2Q/V}$ and energies $E_{th} = (3/4) n_e V k_B T_e$ (Figures 1(e) and (f)). Both quantities vary as $\sqrt{V}$, so are not very sensitive to uncertainties in the volume.

During the SXR peak at ~00:31:30 UT (Figure 2, inset), the centroid positions of the 6.3–7.3 keV, 9–12 keV, and 17–18 keV emission vary linearly (with $\chi^2 < 1$) with the fractional count contribution of the super-hot component (~63%, ~76%, and ~95%, respectively), consistent with spatially distinct super-hot and hot sources whose centroids are separated by ~11.7 ± ~0.7 arcsec. The *RHESSI* imaging data can also be expressed as X-ray visibilities[5] — using the above fractional count percentages, the contributions of the super-hot and hot sources to the total visibilities in each energy band can be computed individually (Caspi 2010). Images created from such visibilities are consistent with two well-separated plasmas, with the super-hot source farther from the footpoints than the hot source throughout the flare. The super-hot source's separation from the footpoints increases over time, and at times, the source is elongated up to ~2× in that direction.

Table 1 gives the parameters of the super-hot and hot plasmas at times during the flare. At the beginning of the flare pre-impulsive phase, we find from the Fe and Fe–Ni lines that the temperature is already ≳25 MK. At the peak of pre-impulsive phase, when faint footpoints are visible, we find that the plasma has become super-hot, and a second hot, but lower temperature, component is also required to fit the data (Figure 4). During the impulsive phase, the super-hot plasma temperature peaks at ~44 MK during the non-thermal HXR (60–100 keV) peak. The super-hot emission measure then is only ~20% of the peak value, reached ~9 minutes later. The super-hot plasma cools rapidly as the HXR emission decreases by a factor of ~10 and drops below 30 MK when the HXR emission nears background, reaching a minimum of ~21 MK during the flare decay. The total super-hot thermal energy, however, decreases relatively slowly, dropping by only a factor of ~4.6 from its maximum until spacecraft nighttime.

The hot plasma begins at ≲18–21 MK at least as early as ~00:25 UT, when footpoints begin to be visible, and peaks at ~24 MK ~1 minute after the super-hot temperature peak, decaying relatively slowly thereafter. The hot plasma is always significantly (~7.5–25 MK) cooler than the super-hot component, and its temperature varies within a much narrower range (~13–24 MK). The best-fit temperature and emission measure of this hot plasma agree closely with those derived from *GOES* — to within ~5%, and ~20%, respectively — except before ~00:38 UT, when the super-hot and non-thermal emission are intense (likely contaminating the *GOES* measurements). The large fluctuations over short timescales (~20–60 s) in the hot temperature and emission measure (Figures 1(c) and (d)), such as around the temperature peak at ~00:29–00:33 UT, are generally anti-correlated and thus likely artifacts of fitting — the hot continuum contributes only ~10%–20% of the total ~3–20 keV counts, and equally acceptable fits to the spectra are achieved when the values are smoothed over 3–5 intervals.

---

[3] http://hesperia.gsfc.nasa.gov/rhessidatacenter/spectroscopy.html
[4] http://www.lmsal.com/solarsoft/
[5] http://sprg.ssl.berkeley.edu/~tohban/wiki/index.php/RHESSI_Visibilities

Compton backscatter of coronal X-rays from the photosphere ("albedo") was neglected because of the flare's ~73° heliocentric angle. Applying an isotropic-source albedo correction (Kontar *et al.* 2006) to the Figure 2 spectrum yields only small (≲10%) changes in both continuum temperatures and the hot emission measure, but a ~33% drop in the super-hot emission measure; however, the super-hot density and energy change by only ~18%, and the derived centroid separation changes by only ~15% (~2σ) to ~9.9 ± ~0.6 arcsec, so our results are not significantly affected.

The CHIANTI-predicted line fluxes and ratio are not significantly affected (≲2%) by the albedo correction, which therefore cannot explain the quantitative disagreement with the observations. The ionization timescales (cf. Jordan 1970; Phillips 2004) for Fe XXV and Ni XXVII — the primary line contributors — at the measured temperatures and densities are generally <1 s and never exceed ~13 s, much shorter than the temperature change timescale, $T_e(\partial T_e/\partial t)^{-1}$, which always exceeds ~130 s, suggesting that ionization equilibrium is always maintained. Phillips *et al.* (2006) suggested a potential inaccuracy in CHIANTI's Fe ionization fractions that may account for the systematic discrepancies.

At the peak of the pre-impulsive phase (Figure 4), the measured Fe and Fe–Ni line fluxes constrain the super-hot plasma temperature to be between ~29 and ~37 MK. The low-energy cutoff of the cospatial non-thermal source must then be below ~20 and ~27 keV, respectively. At the measured thermal density of ~$2\times10^{11}$ cm$^{-3}$, the collisional energy loss time (Lin 1974) for 20–100 keV non-thermal electrons is only ~0.03 to ~0.3 seconds. Assuming a thick-target model (Brown 1971), the total non-thermal energy deposition during the entire pre-impulsive period is ≳$1.0\times10^{31}$ and ≳$2.4\times10^{30}$ erg, respectively, less than the lower limit of ~$1.7\times10^{31}$ erg estimated by Holman *et al.* (2003) and significantly smaller than the upper limit of ~$4\times10^{32}$ erg (cf. Lin *et al.* 2003).

## 3. DISCUSSION

The simplest physical model that fits the *RHESSI* images and spectra throughout the 2002 July 23 X4.8 flare consists of two distinct thermal components: a super-hot plasma high in the corona, and the hot, ~10–20 MK plasma normally detected by *GOES* closer to the footpoints. Preliminary analysis of a second flare — the 2003 November 02 X8.1 event — at the HXR peak also reveals two distinct plasmas, with temperatures of ~45 MK and ~18 MK. A similar double-temperature behavior was reported for two other flares using combined observations from the Bent Crystal Spectrometer (BCS) and the Hard X-ray Imaging Spectrometer (HXIS) on *Solar Maximum Mission* (Jakimiec *et al.* 1988). We note that while our spectral model assumed two discrete temperatures, preliminary fits to the July 23 continuum spectra using a continuous differential emission measure model also yielded a sharply-peaked bimodal distribution (McTiernan & Caspi 2010).

The energy densities (assuming $T_i=T_e$) of the July 23 super-hot and hot plasmas at the time of the peak super-hot temperature (~00:28:30 UT) are ~4800 and ~5900 erg cm$^{-3}$, respectively. To magnetically confine these plasmas, coronal field strengths exceeding ~350 and ~380 G, respectively, are required. A preliminary survey of 37 large flares revealed that every super-hot X-class flare (11 of 12 surveyed) required field strengths exceeding ~220–460 G, while cooler M-class flares that did not reach super-hot temperatures (23 of 25 surveyed) required fields of ~60–400 G, suggesting that strong coronal fields are necessary (but not sufficient) for achieving temperatures above ~30 MK (Caspi 2010).

At the beginning of the July 23 pre-impulsive phase, a hot thermal plasma ($T \gtrsim 25$ MK, $n_e \gtrsim 10^{10}$ cm$^{-3}$) is already present in the corona, cospatial with a non-thermal coronal HXR source. There is little or no detectable HXR footpoint emission, however, indicating that chromospheric evaporation is negligible and suggesting a coronal origin for the super-hot plasma.

In the standard flare reconnection model (e.g., Sturrock 1968; Shibata 1996), plasma within the reconnection region is energized directly by Joule heating. As the reconnected, elongated magnetic field relaxes to a more potential configuration, the field lines shorten and the field strength increases, causing the hot plasma to be further energized by Fermi acceleration and betatron heating, respectively. An order-of-magnitude estimate (Caspi 2010) suggests that the resulting temperatures and densities are consistent with those observed for the super-hot plasma during the onset of the July 23 flare.

The separate, denser, lower-altitude *GOES*-temperature plasma becomes distinct only when the HXR footpoint emission becomes significant (~00:25 UT), as expected if the cooler plasma originates from evaporation of chromospheric material by the impacting accelerated electrons.

Thus, the *RHESSI* spectra and images show that throughout the 2002 July 23 X4.8 flare, the super-hot thermal plasma is distinct, both spectrally and spatially, from the commonly observed hot plasma. The super-hot source location and its presence at the flare onset — in the absence of HXR footpoint emission — point to a coronal origin for the super-hot plasma, while the hot plasma originates primarily from chromospheric evaporation. The super-hot and hot plasmas thus arise from fundamentally different physical processes, and this bimodal behavior appears common amongst super-hot flares.

This work was supported in part by NASA contract NAS5-98033. A. Caspi was also supported in part by NASA grant NNX08AJ18G, and R. Lin by the WCU grant (No. R31-10016) funded by the Korean Ministry of Education, Science and Technology. The authors would like to thank B. Dennis, H. Hudson, G. Hurford, S. Krucker, J. McTiernan, K. Phillips, R. Schwartz, and D. Smith for helpful discussions.

**Table 1**
Super-hot and Hot Plasma Parameters During the 2002 July 23 X4.8 Flare

| Time[a] (UT) | | Super-hot Plasma[b] | | | Hot Plasma[b] | | |
|---|---|---|---|---|---|---|---|
| | | $T_e$ (MK) | EM ($10^{49}$ cm$^{-3}$) | Energy ($10^{28}$ erg) | $T_e$ (MK) | EM ($10^{49}$ cm$^{-3}$) | Energy ($10^{28}$ erg) |
| *Pre-impulsive phase:*[c] | | | | | | | |
| Start[d] | ~00:22:00 | ~23 | ~0.032 | ~5.5 | — | — | — |
| Peak[e] | ~00:25:40 | ~33 | ~0.082 | ~15 | ~19 | ~0.45 | ~20 |
| *Impulsive phase:* | | | | | | | |
| HXR peak | ~00:28:30 | ~44 | ~0.86 | ~41 | ~24 | ~4.2 | ~50 |
| Super-hot EM peak | ~00:37:30 | ~32 | ~4.4 | ~57 | ~18 | ~23 | ~72 |
| HXR at background | ~00:42:00 | ~30 | ~2.2 | ~65 | ~19 | ~13 | ~100 |
| *Decay phase:* | | | | | | | |
| Super-hot T minimum | ~01:06:18 | ~21 | ~0.81 | ~40 | ~14 | ~6.6 | ~73 |
| Just prior to eclipse | ~01:14:58 | ~24 | ~0.13 | ~15 | ~14 | ~3.3 | ~44 |

**Notes.**
[a] Quoted times represent the center of the interval used for spectral analysis.
[b] Quoted values refer to electrons.
[c] During the pre-impulsive phase, the Fe and Fe–Ni line observations constrain the super-hot temperature and emission measure, providing upper and lower limits.
[d] At the start of the pre-impulsive phase, the simplest acceptable model is a single component, although the line observations cannot exclude two components with widely-separated temperatures.
[e] During the peak of the pre-impulsive phase, the quoted values represent the midpoints of the ranges allowed by the Fe and Fe–Ni line constraints.

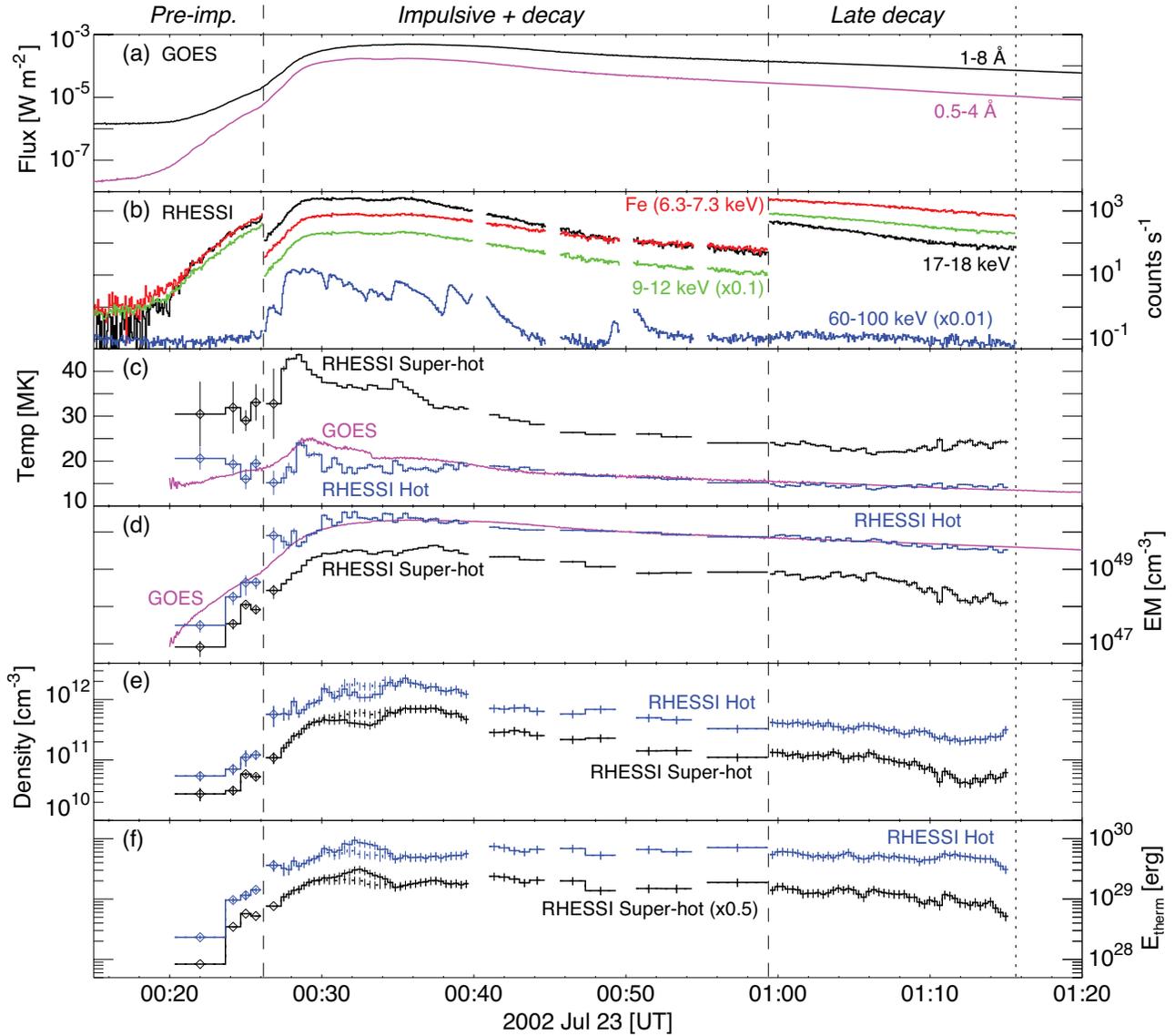

**Figure 1.** (a) Temporal evolution of the GOES fluxes; (b) RHESSI count rates — dashed vertical lines indicate attenuator-state transitions; (c) super-hot (black), hot (blue), and GOES (magenta) plasma temperatures; (d) emission measures; (e) super-hot (black) and hot (blue) electron densities; (f) total energies for the 2002 July 23 X4.8 flare. Early on (~00:20–00:27 UT), the diamonds with error bars repre-sent the constraints derived from the Fe and Fe-Ni line measurements. The dotted lines in (e) and (f) represent a correction for an unusual variation in the source volume during ~00:30–00:35 UT, when two small, spatially-distinct sources appear to be simultaneously bright (see Caspi 2010 for details).

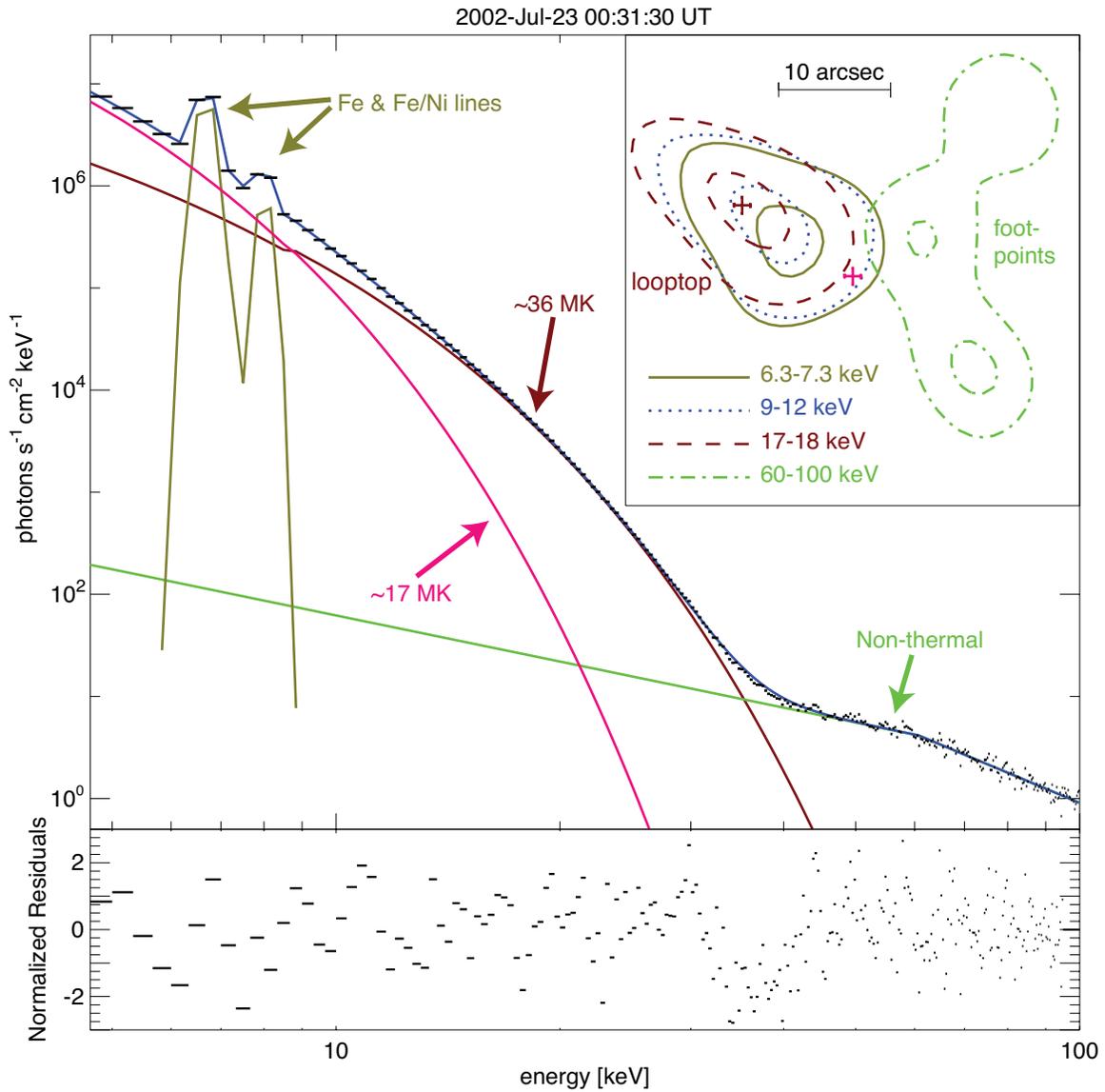

**Figure 2.** Photon flux spectra (black), model fit (Fe and Fe-Ni lines: olive; super-hot: brown; hot: magenta; non-thermal: green; total model: blue), and normalized residuals during the RHESSI SXR peak (~00:31:30 UT), when the super-hot component is strongest. Inset: 50% and 90% contours of 6.3–7.3 (olive solid), 9–12 (blue dotted), 17–18 (brown dashed), and 60–100 keV (green dot-dashed) images at the same time; the crosses denote the derived centroid locations (and uncertainties) of the super-hot (brown; left) and hot (magenta; right) components.

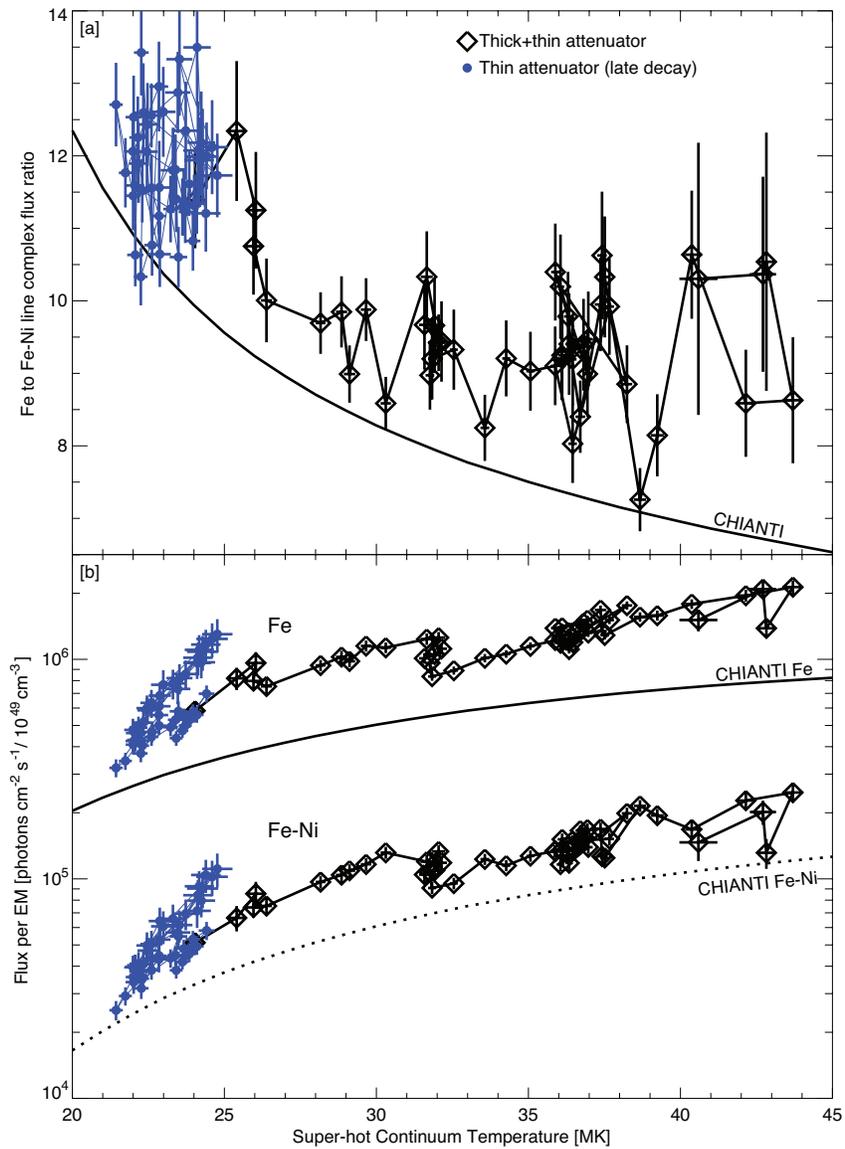

**Figure 3.** (a) Measured Fe to Fe-Ni line flux ratios and (b) individual line fluxes, normalized by the super-hot component emission measure, vs. measured super-hot continuum temperature during the impulsive and decay phases. The black curves show the CHIANTI predictions for an isothermal plasma; the systematically larger observed ratios and fluxes are expected from the contribution of the second, hot isothermal component to the lines. However, accounting for the hot isothermal yields predicted ratios and fluxes that significantly exceed the observed values (Caspi 2010).

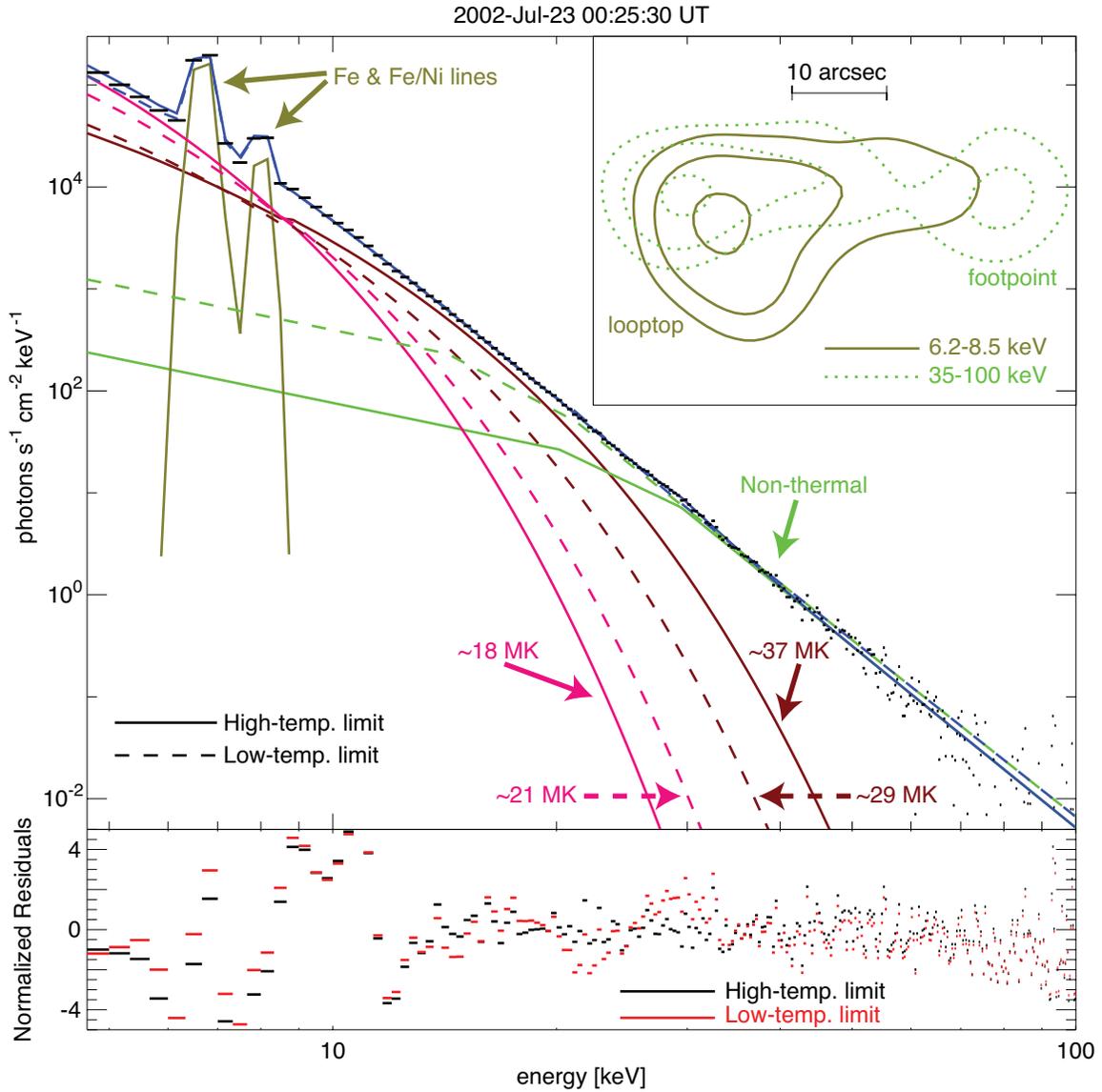

**Figure 4.** Photon flux spectra (black) during the peak of the pre-impulsive phase (~00:25:40 UT), with two acceptable model fits showing the upper (solid) and lower (dashed) temperature limits of the super-hot component (brown), as constrained by the Fe and Fe-Ni lines (olive). Hot isothermal (magenta) and non-thermal (green) components are also required. Inset: 30%, 50%, and 90% contours of 6.2–8.5 (olive solid) and 35–100 keV (green dotted) images at the same time. The peak non-thermal emission appears to be above the thermal looptop; the faint footpoint contains only ~16% of the total non-thermal flux within the 50% contour (~29% within the 30% contour).